\documentclass[12pt]{article}
\begin{document}
\begin{center}
\large{\bf{The Signals and Systems Approach to Quantum Computation}}\\
~\\
H. Gopalkrishna Gadiyar $^{a}$, K M Sangeeta Maini $^{a}$,\\ R. Padma $^{a}$, H. S. Sharatchandra$^{b}$
\end{center}

\begin{flushleft}
$^{a}$ AU-KBC Research Centre,
M. I. T. Campus, Anna University,\\
~~~Chromepet
Chennai 600 044 INDIA\\
$^{b}$ Institute of Mathematical Sciences,
C. I. T. Campus, Taramani P. O.,\\
~~~Chennai 600 113 INDIA\\
~~\\
$^{~}$ E-mail:
gadiyar@au-kbc.org, sangeeta\_maini@au-kbc.org, \\~~~~~~~~~~~~~padma@au-kbc.org, sharat@imsc.res.in
\end{flushleft}

\vspace{1.5cm}

\noindent{\bf{Abstract}}

In this note we point out the fact that the proper conceptual setting of quantum computation is the theory of Linear Time Invariant systems. To convince readers of the utility of the approach, we introduce a new model of computation based on the orthogonal group. This makes the link to traditional electronics engineering clear. We conjecture that the speed up achieved in quantum computation is at the cost of increased circuit complexity.

\newpage

This paper is an extension of the ideas of Pravin Varaiya and Edward A. Lee, University of California, Berkeley ([1],[2]), who have tried to unify electrical engineering and computer science curricula. Traditionally, computer science and electrical engineering did not have much in common. However, the rise of Internet has led to a virtual merger in the industry of telecom and computers. This has forced a rethink of introductory courses which is what the two authors have done. They have essentially called for a merger of automata theory with the traditional signals and systems approach.  

Recently there has been a lot of interest in quantum computation following the results of Shor [3] and Grover ([4],[5]). The discovery of factoring and search algorithms which are computationally efficient has led to an interest in experimental implementation of quantum computers. Historically quantum computation has been studied by theoretical physicists and computer scientists. Physicists have focussed on the quantum aspects of quantum computation and computer scientists towards designing efficient algorithms and classifying problems into new complexity classes. Several issues are discussed simultaneously which lead to a lot of confusion for someone who wishes to make a table-top quantum computer. Thus there is talk of superposition, interference, entanglement, non-determinism, non-clonability and non-locality. All this leads to the feeling that quantum computers are exotic devices.

In this note we apply Occam's razor to cut through the confusion and make the subject accessible to experimentalists and electronics engineers.

In the so-called Berkeley-approach, electronics and computer science are treated in a very general framework of signals and systems. Linear Time Invariant (LTI) systems are a special case of the Berkeley approach and more general systems can also fit in. We wish to emphasize that quantum computers fit into the framework of LTI systems. It is well known that universal gates for quantum computation have been identified. Hence, any implementation of these gates would lead to construction of quantum computers.

We do not take the more general approach of ([1],[2]). For our purposes it is enough to have the framework of LTI systems. In [6] Stedman fits electronics and quantum mechanics into the same framework - a fact well known to many people.

Let us denote the input signals by $x_i(t)$ and the corresponding output signals by $y_i(t)$. If $f$ is the mathematical effect of an identifiable device (the system), then the system possesses the following properties. \\ 
~\\
(a)~Linearity which means that the principle of superposition holds, that is, if 
\begin{equation}
y_1(t)~=~f(x_1(t))
\end{equation}\\
and 
\begin{equation}
y_2(t)~=~f(x_2(t))
\end{equation}\\
then for linear systems \\
\begin{equation}
a_1~y_1(t) + a_2~y_2(t)~ =~ f(a_1~x_1(t)+ a_2~x_2(t)).
\end{equation}\\
(b)~Time invariance essentially means that we get the same output if we repeat an experiment with the same input except for a time delay. Mathematically,\\
\begin{equation}
y(t-\tau)~=~f(x(t-\tau)).
\end{equation}\\
(c)~Causality or realizability means that if the input signal $x(t)$ lies in the interval $0<t<\infty$, then the output signal 
\begin{equation}
y(t)~=~f(x(t)) 
\end{equation}\\
also lies in the same interval $0<t<\infty$,
that is, the effect (the output) cannot precede the cause (the input).

We therefore see a possibility of exploiting this mathematical and physical commonality to design and implement quantum computers in new settings. This liberates us from thinking about implementations which are necessarily atomic or quantum. In [7] Merkle discusses the possibility of implementing reversible computation using conventional technologies. The experimentalists have essentially tried to use a variety of techniques but all of these are essentially various physical manifestations of the same underlying mathematical structure of two-level systems [8]. For example,\\
\begin{equation}
i\hbar\frac{d}{dt}|\Psi>~=~{H}|\Psi>
\end{equation}
would be
\begin{equation}
i\hbar\frac{d}{dt}
\left[ \begin{array}{c}
  C_1(t) \\
  C_2(t) \\
\end{array} \right]
~=~
\left[ \begin{array}{cc}
  H_{11}	& H_{12} \\
  H_{21}	& H_{22} \\
\end{array} \right]
\left[ \begin{array}{c}
  C_1(t) \\
  C_2(t) \\
\end{array} \right]
\end{equation}
It is well known [9] that these differential equations are LTI systems. In this form the close relationship to analog electronics, where differentiators and integrators are routinely built, is seen. We do not get into questions of measurement by appealing to the principle of deferred measurement. This states that we can push all measurement operations to the last stage [10].

To convince readers that this approach can yield new results we introduce a new class of computers which we call Orthogonal computers and their basic unit becomes Orthogonal bit (O-bit). Here we recall the well known fact that the bit is the fundamental unit of a classical computer. A quantum computer too has the bit as its fundamental concept but it is quantum in nature and therefore this quantum bit is called `qubit'. The special thing about this qubit is that, unlike the classical bit, it can be in a superposition of states. For example, a classical bit exists either in state 0 or state 1 but a qubit can, apart from this, exist very much in a state, which is something in between state 0 and state 1. This requires a standard notation, which is due to Dirac. The Dirac notation deals with $< |$ and $| >$ called the `Bra' and the `Ket' (derived from Bra-c-Ket). Thus, in this notation the states 0 and 1 are written as $|0>$ and $|1>$ respectively and the `in between state' can be understood from the state of the system represented as follows:
\begin{equation}
                  |\psi>~=~\alpha |0> + \beta |1> 
\end{equation}
Here $\alpha$ and $\beta$ are complex numbers. This means that the state of a system is represented by the state vector $|\psi>$ (in the Ket notation) and is the linear combination of states or superposition of states. It can exist in the state $|0>$ with probability $|\alpha|^2$ and in the state $|1>$ with probability $|\beta|^2$. Since probabilities sum to one, 
\begin{equation}             
|\alpha|^2 + |\beta|^2~=~1  \, .           
\end{equation}\\
Also the state of a qubit is a two-dimensional complex vector space. The states $|0>$ and $|1>$ are known as computational basis states, and form an orthonormal basis for this vector space. 

As in the earlier case we represent the O-bits as $|0)$ and $|1)$ (Like $<|$ and $|>$ are called Bra- and -Ket, we wish to call $(~|$ and $|~)$ as Smi- and -Ley as we hope that this version of computers will be more palatable to hackers.) The state of a system, represented by $|V)$, would be real linear superposition of the O-bits as follows:
\begin{equation}
|V) = a|0) + b|1)
\end{equation}
where $a$ and $b$ are real, satisfying
\begin{equation}
 a^2 + b^2 = 1 \, .
\end{equation}

To make an orthogonal computer in a lab we would have to represent
$|0)$ and $|1)$ as signals which should satisfy the following properties: 
\begin{equation}
{(0|0)=1,~(1|1)=1,~(0|1)=0,~(1|0)=0.}
\end{equation} 
This can be practically implemented by mapping the O-bits
\begin{equation}
|0)~\rightarrow~f_{0}(\xi)=\left \{ \begin{array}{l} \sin\xi ,~ \rm{if}~0\leq\xi\leq2\pi\\
0, ~ \rm{elsewhere}
\end{array} \right.
\end{equation}
\begin{equation}
|1)~\rightarrow~f_{1}(\xi)=\left \{ \begin{array}{l} \cos\xi ,~ \rm{if}~0\leq\xi\leq2\pi\\
0, ~ \rm{elsewhere}
\end{array} \right.
\end{equation}
and the inner product\\
\begin{equation}
(0|0)\rightarrow~\frac{1}{\pi}\int_{0}^{2\pi}\sin^2\xi~d\xi
\end{equation} 
\begin{equation}
(1|1)\rightarrow~\frac{1}{\pi}\int_{0}^{2\pi}\cos^2\xi~d\xi
\end{equation} 
\begin{equation}
(0|1)\rightarrow~\frac{1}{\pi}\int_{0}^{2\pi}\sin\xi~\cos\xi~d\xi
\end{equation} 
\begin{equation}
(1|0)\rightarrow~\frac{1}{\pi}\int_{0}^{2\pi}\cos\xi~\sin\xi~d\xi
\end{equation} 
where $\xi$ could be either in the frequency domain or in the time domain. It is immediate that orthogonal computers can be implemented using usual electronics.

Questions about the construction of universal gates, analysis of complexity classes and construction of algorithms for orthogonal computers will be done in a subsequent paper.

Next, we conjecture that there is a simple trade-off of resources: the faster the computation the more complex the circuits become. This can be seen by an analysis of the fast Fourier transform and quantum Fourier transform. If we recast the quantum Fourier transform in co-ordinate representation, it becomes intimately related to the fast Fourier transform [11].

Let $x_0,...,x_{N-1}$ and $y_0,...,y_{N-1}$ be complex numbers where 
\begin{equation}
y_k~\stackrel{\mathrm{def}}{=}~\frac{1}{\sqrt{N}}\sum_{j=0}^{N-1} x_j~e^{\frac{2{\pi}ijk}{N}}
\end{equation}
is the discrete Fourier transform of $x_j$.
The quantum Fourier transform on an orthonormal basis $|0>,.....,|N-1>$ is defined to be a linear operator with the following action on the basis states,
\begin{equation}
|j>~\rightarrow~\frac{1}{\sqrt{N}}\sum_{k=0}^{N-1} e^{\frac{2{\pi}ijk}{N}}|k> \, .
\end{equation}
In other words, the action on an arbitrary state may be written as
\begin{equation}
\sum_{j=0}^{N-1}x_j|j>~\rightarrow~\sum_{k=0}^{N-1}y_k|k>
\end{equation}
where the amplitudes $y_k$ are the discrete Fourier transform of the amplitudes $x_j$. This transformation is unitary and is implementable.
The product representation of the discrete Fourier transform is given by
\begin{equation}
|j_1,...,j_n> \rightarrow
\frac{(|0>_1~+~e^{2{\pi}i0.j_n}|1>_1)...(|0>_n~+~e^{2{\pi}i0.j_1j_2...j_n}|1>_n)}{2^{\frac{n}{2}}}.
\end{equation}
In the co-ordinate form
\begin{equation}
y_{i_1...i_n}~=~\sum_{j_1}...\sum_{j_n} ~U_{i_1j_1}^1~U_{i_2j_2}^2...U_{i_nj_n}^n~x_{j_1...j_n} \, ,
\end{equation}
where the $U_{i_{\alpha}j_{\alpha}}^{\alpha}$ are matrices obtained by taking the inner product of Bra vectors $<i_1...i_n|$ with equation (22).
It then becomes clear that this is a fast Fourier transform where $x$ and $y$ are represented as binary numbers.

It can be seen that in co-ordinate representation we are implementing the Fourier transform in a tensor form. Each tensor co-ordinate is transformed by a phase factor but the complexity of the circuit increases rapidly. We therefore conjecture that the speed-up in quantum computation is due to a trade-off between speed and circuit complexity. 

\newpage
\noindent{\bf{Acknowledgements}}

H.Gopalkrishna Gadiyar would like to thank Dr. L. Chander (currently at Infosys Technologies Limited, Bangalore) for discussions in early stages of this work. He also wishes to thank his colleague Dr. S. Srikanth for pointing out the Berkeley approach ([1], [2]). K M Sangeeta Maini would like to thank Professor R. Simon, Institute of Mathematical Sciences, Chennai for an invitation to the meeting on Quantum Computation and Quantum Information sponsored by L.A. Meera Memorial Trust, Mysore in 2002. 

\vspace{0.5cm}
\noindent{\bf{References}}

\begin{enumerate}
\item[1] Edward A. Lee and Pravin Varaiya, Introducing signals and systems - The Berkeley approach, Proc. SPE Workshop, Hunt, Texas, October 15 - 18, 2000.
\item[2] Edward A. Lee and Pravin Varaiya, Designing a relevant lab for introductory signals and systems, Proc. SPE Workshop, Hunt, Texas, October 15 - 18, 2000.
\item[3] P.W. Shor, Algorithms for Quantum Computation: Discrete Logarithms and Factoring, Proceedings, 35th Annual Symposium on Foundations of Computer Science, IEEE, New York (1994), pp. 124­134
\item[4] L.K. Grover, Quantum Mechanics helps in searching for a needle in a haystack, Phys.Rev.Lett., 79, 325-8.
\item[5] L.K. Grover, Quantum computers can search rapidly by using almost any transformation, Phys.Rev.Lett., 80, 4329-32.
\item[6] G.E. Stedman, Contemp. Phys., 1968, vol. 9, No. 1, 49-69.
\item[7] R.C. Merkle, Nanotechnology, 4(1993), 21-40 (See discussion in M. Li and P. Vit$\acute{a}$nyi, An Introduction to Kolmogorov Complexity and Its Applications, Springer - Verlag, 1993.)
\item[8] R.P. Feynman, R.B. Leighton, M. Sands, The Feynman Lectures on Physics, vol. III, Addison-Wesley, 1965 (Indian edition, Narosa Publications).
\item[9] Alan V. Oppenheim, Alan S. Willsky, S.Hamid Nawab, Signals and Systems, Prentice-Hall,Inc.U.S.A. (Indian edition, Prentice-Hall, India, 1999).
\item[10] M.A. Nielson, I.L. Chuang, Quantum Computation and Quantum Information, Cambridge University Press, 2000.
\item[11] E. Horowitz, S. Sahini, S. Rajasekaran, Fundamentals of Computer Algorithm, 1998, W. H. Freeman and Co. (Indian edition, Galgotia Publications).
\end{enumerate}
\end{document}